\documentclass[aps,prb,twocolumn,showpacs,superscriptaddress]{revtex4-1}

\usepackage{amsmath}
\usepackage{amssymb}
\usepackage{graphicx}
\usepackage{bm}

\usepackage{color}

\begin{document}

\title{Supersymmetry in the Majorana Cooper-Pair Box}
\author{Jascha Ulrich}
\affiliation{JARA-Institute for Quantum Information, RWTH Aachen University,
	D-52074 Aachen, Germany}
\author{\.{I}nan\c{c} Adagideli}
\affiliation{Faculty of Engineering and Natural Sciences, Sabanci University, Orhanli-Tuzla, Istanbul, Turkey}
\author{Dirk Schuricht}
\affiliation{Institute for Theoretical Physics, Center for Extreme Matter and
	Emergent Phenomena, Utrecht University, Leuvenlaan 4, 3584 CE Utrecht,
	The Netherlands}
\author{Fabian Hassler}
\affiliation{JARA-Institute for Quantum Information, RWTH Aachen University,
	D-52074 Aachen, Germany}

\pacs{%
  85.25.Cp, 
  11.30.Pb, 
  73.23.-b, 
  74.50.+r  
}

\date{May 2014}

\begin{abstract}
Over the years, supersymmetric quantum mechanics has evolved from a toy model
of high energy physics to a field of its own.  Although various examples of
supersymmetric quantum mechanics have been found, systems that have a natural
realization are scarce.  Here, we show that the extension of the
conventional Cooper-pair box by a $4\pi$-periodic Majorana-Josephson coupling
realizes supersymmetry for certain values of the ratio between the
conventional Josephson and the Majorana-Josephson coupling strength.  The
supersymmetry we find is a ``hidden'' minimally bosonized supersymmetry that
provides a non-trivial generalization of the supersymmetry of the free
particle and relies crucially on the presence of an anomalous Josephson
junction in the system.  We show that the resulting degeneracy of the energy
levels can be probed directly in a tunneling experiment and discuss the
various transport signatures.  An observation of the predicted level
degeneracy would provide clear evidence for the presence of the anomalous
Josephson coupling.
\end{abstract}

\maketitle

\section{Introduction}

Supersymmetric quantum mechanics in its conventional form describes systems
that consist of two sectors (dubbed ``fermionic'' and ``bosonic'' sector)
where, apart from the ground state, each state has a partner state at equal energy
in the other sector.\cite{witten:81, cooper:83} Thus, the supersymmetry leads to a
degeneracy of the eigenstates that cannot be explained in the conventional
framework of continuous symmetries and their respective higher-dimensional
irreducible representations.  In the condensed matter context, the main focus
has been on a technical usage of supersymmetric quantum mechanics allowing,
for example, the algebraic construction of the spectra of non-supersymmetric
Hamiltonians\cite{cooper:95} or partially analytic approaches to
(supersymmetric) lattice models like the ferromagnetic $t-J$
model\cite{fendley:03b}, the XXZ
chain\cite{fendley:03, beccaria:05} or quantum-critical systems\cite{huijse:08}.  The
supersymmetries discussed in these works do not necessarily involve a
degeneracy of the eigenstates on a physical level as the authors invoke
supersymmetry for providing an additional structure which helps in understanding the (exact)
solution of the problem. 
Level degeneracies due to a supersymmetry have been
investigated in the context of the hydrogen atom\cite{tangerman:93} and their
experimental signatures have been discussed for cold gases implementing high-energy physics 
inspired models\cite{snoek:05, yu:08}.  In
this paper, we show that adding a Majorana-Josephson junction of the right
coupling strength to a Cooper-pair box leads to a degeneracy of all excited
energy levels due to supersymmetry.  The supersymmetry we find is a ``hidden''
minimally bosonized supersymmetry\cite{plyushchay:96} and realizes a
non-trivial generalization of the supersymmetry of the free particle in one
dimension\cite{rau:04} to the presence of a potential.  Moreover, we show that
the supersymmetry can be directly probed in a tunneling experiment giving
access to spectral properties of the system.

\begin{figure}[tbp]
\includegraphics[width=0.4\textwidth]{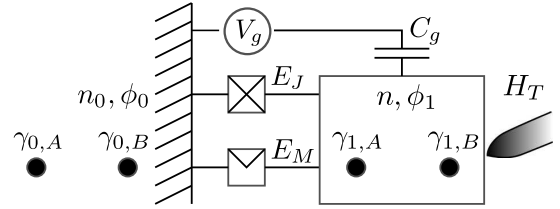}
\caption{Setup corresponding to the model Hamiltonian
Eq.~\eqref{eq:ham_system_general}, describing a superconducting island with
charge $n$, superconducting phase $\phi_1$ and Majorana bound states
$\gamma_{1,A/B}$ coupled to a ground superconductor with charge $n_0$,
superconducting phase $\phi_0$ and Majorana bound states $\gamma_{0,A/B}$.
The coupling is realized through a Josephson junction (depicted by a boxed
cross) with Josephson energy $E_J$ and a Majorana-Josephson junction (depicted
as a half-cross) between bound states $\gamma_{0,B}$ and $\gamma_{1,A}$ with
coupling strength $E_M$.  The island is coupled via a capacitance $C_g$ to a
gate voltage $V_g$.  The tunnel Hamiltonian $H_T$ denotes the possibility to
couple the bound state $\gamma_{1,B}$ to a normalconducting tunneling
tip.}
\label{fig:system_setup}
\end{figure}

Majorana fermions have attracted a lot of attention in the last years
\cite{alicea:12, beenakker:13} due to their potential for quantum
computation \cite{nayak:08}.  At first sight, superconducting systems hosting
Majorana fermions appear to be prime candidates for the realization of
supersymmetry since they involve both bosonic (Cooper-pair condensate) and
fermionic (Majorana fermions) degrees of freedom.  Indeed, a supersymmetry in
space-time has recently been shown to arise at an interface of two topological
superconductors in two dimensions \cite{tsvelik:12} as well as at the quantum
phase transition between a trivial and a topological superconductor in
arbitrary dimensions \cite{grover:13}. In contrast, here we want to focus on a
realization of supersymmetric quantum mechanics and its associated level
degeneracies with the help of Majorana
fermions that does not rely on their fermionic properties, but only on the
simultaneous presence of an anomalous $4\pi$-periodic Josephson coupling and a
normal one.

In Sec.~\ref{sec:majorana_cpb}, we introduce our system of interest, the
Majorana Cooper-pair box.  In Sec.~\ref{sec:susy_qm_cpb}, we give a short
outline of supersymmetric quantum mechanics and show that the Majorana
Cooper-pair box is supersymmetric for a certain ratio of Josephson to
Majorana-Josephson coupling strength.  In Sec.~\ref{sec:tunneling_current}, we
discuss a tunneling experiment for a direct probe of the level degeneracy
predicted by the supersymmetry before we conclude by summarizing our main
findings and discussing possible experimental realizations.

\section{Majorana Cooper-pair box}
\label{sec:majorana_cpb}

The system of interest is depicted in Fig.~\ref{fig:system_setup}.  It is
based on the well-known Cooper-pair box, which consists of a superconducting
island (with phase $\phi_1$) that is coupled to a ground superconductor (with
phase $\phi_0$) via a gate voltage $V_g$ with capacitance $C_g$ and a
Josephson junction with Josephson energy $E_J$.  The Cooper-pair box is
extended by a $4\pi$-periodic Josephson junction with coupling strength $E_M$
which is characteristic for topological superconductors.  The $4\pi$-periodic
Josephson effect comes along with one Majorana zero mode denoted by
$\gamma_{0,B}$ and $\gamma_{1,A}$ on either side of the junction.  Exchange of
single electrons leads to the hybridization energy $i E_M \gamma_{0,B}
\gamma_{1,A} \cos(\phi/2)$ where $\phi =\phi_1 - \phi_0$ is the
superconducting phase difference.  Due to topological constraints, there are
always an even number of Majorana zero modes on each superconducting island.
Thus, we have to take into account two additional Majorana bound states
$\gamma_{0,A}$ and $\gamma_{1,B}$.  We assume that the Majorana modes on the
same superconductor are sufficiently separated such that we can neglect the
exponentially small energy splitting.  In a specific realization of our
proposed system, the Majorana bound states could be hosted, for example, at
the ends of semiconductor nanowires placed on top of a conventional s-wave
superconductor.\cite{oreg:10, lutchyn:10} However, our discussion is
independent from the specific way the Majorana bound states are realized.

The total Hamiltonian of the system reads
\begin{align}\label{eq:ham_system_general}
H_\gamma = E_C (n - n_g)^2 + E_J \big(1 - \cos\phi) \nonumber \\ 
+  i E_M
\gamma_{0,B} \gamma_{1,A} \cos(\phi/2);
\end{align}
the last term is the Majorana Josephson coupling explained in details above.
The first term is associated with the electrostatic charging energy $E_C =
e^2/2C_g$ of having $n$ electrons on the superconducting island and $n_g = C_g
V_g/ e$ is the preferred electron number (with $e>0$ the elementary charge) on
the island set by the gate voltage. The second term proportional to $E_J =
\hbar I_c/2e$ arises due to the conventional Josephson coupling exchanging
Cooper-pairs with a critical current $I_c$. In deriving the Hamiltonian, we
have assumed a large ground superconductor such that there is no charging 
energy associated with it. As a result, the superconducting phase $\phi_0$ of 
the ground superconductor has no dynamics and we can choose a gauge with 
$\phi_0 = 0$. 
The number of electrons $n \in \mathbb{Z}$ and the phase $\phi_1 = \phi$ of 
the superconducting island are conjugate variables and obey the 
angular-momentum algebra
\begin{align}\label{eq:charge_phase_algebra}
[n, e^{\pm i \phi/2}] = \pm  e^{\pm i \phi/2},
\end{align}
such that $e^{\pm i \phi/2}$ corresponds to addition/removal of a single
electron.  The Majorana operators obey the Clifford algebra
\begin{align}
\{\gamma_k, \gamma_l\} = \gamma_k \gamma_l + \gamma_l \gamma_k =2
\delta_{kl}.
\end{align}

Assuming that the temperature is below the superconducting gap and that apart
from the Majorana modes there are no additional Andreev states, an occupation
of the (non-local) fermionic mode spanned by the Majorana bound states
$\gamma_{1,A}, \gamma_{1,B}$ must correspond to the presence of an odd number of electrons on
the superconducting island. Consequently, we have the fermion parity
constraint\cite{fu:10}
\begin{align}\label{eq:fermion_parity_constraint} i \gamma_{1,A} \gamma_{1,B}
	= (-1)^{n}  \end{align}
for the island and an analogous constraint for the ground superconductor.
For each superconductor with a pair of Majorana bound states, the fermion
parity constraint reduces the Hilbert space dimension by a factor of two,
and consequently, the Hilbert space of the 
system~\eqref{eq:ham_system_general} is four times smaller than one would naively
expect.
Since the Majorana degrees of
freedom are slaved to the number operator $n$, they can be explicitly removed via
a unitary transformation $U$, see App.~\ref{sec:app_bosonization_ham} for
details.\cite{heck:11,zazunov:11} We obtain
\begin{align}\label{eq:ham_system} H = E_C (n - n_g)^2 + E_J (1-\cos \phi) +
	E_M \cos(\phi/2) , \end{align}
where $H$ is the projection of the transformed Hamiltonian $U H_\gamma U^\dag$ 
onto the constraint surface. As the Hamiltonian is $4\pi$-periodic in $\phi$,
the charge offsets $n_g$ are only defined modulo 1.  Thus, we restrict
ourselves to $n_g \in [-\tfrac12 , \tfrac12]$ and adjust $n$ accordingly.  For
the important situation with $n_g=0$, $n$ simply counts the number of excess
charges.  We will show that at this particular point the Hamiltonian
\eqref{eq:ham_system} is supersymmetric with a level degeneracy due to the
symmetry provided that $E_M = \sqrt{2 E_J E_C}$.

\section{Supersymmetry}
\label{sec:susy_qm_cpb}
\begin{figure*}[tbp]
\centering
\includegraphics[width=1.\textwidth]{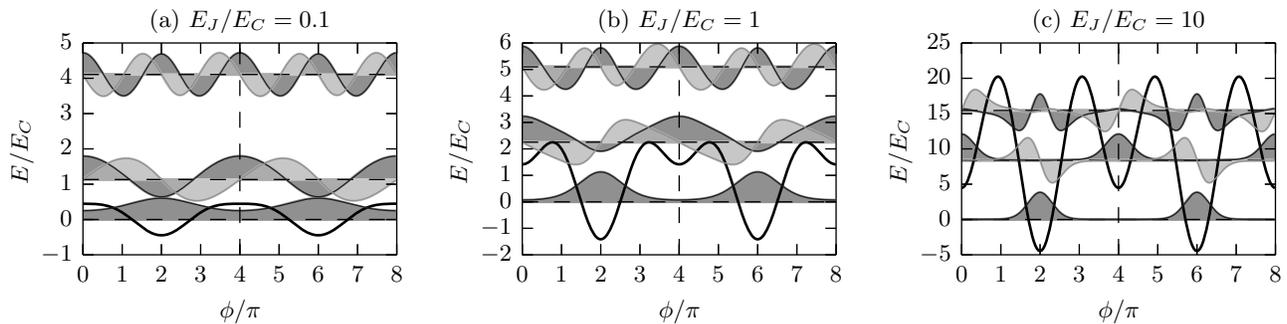}
\caption{Numerically calculated wave functions $\psi_{n,\pm}(\phi)$ of the
Hamiltonian Eq.~\eqref{eq:ham_system} for $n = 0, 1, 2$ at the supersymmetric
point $n_g = 0$, $E_M = \sqrt{2 E_J E_C}$ for different values of $E_J/E_C$
(and thus $\alpha = \sqrt{2 E_J/E_C}$).  The wave
functions are chosen real and are aligned at their corresponding eigenenergies.
States $\psi_{n,+}(\phi)$ in the even parity sector of the superconducting 
phase are plotted in dark gray while the states $\psi_{n,-}(\phi)$ in the odd 
parity sector are plotted in light gray.  The black line represents the 
underlying potential. Both the potential and the states are $4\pi$ periodic. 
We note that due to the supersymmetry, all levels with $n > 0$ are doubly 
degenerate.}
\label{fig:susy_wf}
\end{figure*}
A supersymmetric Hamiltonian $H_Q$ decomposes into a direct sum of two terms
that share the same spectrum up to a possibly missing ground state.  The
structure behind the $N = 1$ supersymmetry in quantum mechanics is generated
by a Hermitian supercharge $Q$ and Hermitian operator $K$ squaring to 1 which
distinguishes the ``bosonic'' and ``fermionic'' sectors.\cite{combescure:04}
In particular, these operators implement the algebra
\begin{align}
 \{Q,Q\}  &= 2H_Q, & \{Q, K\} &= 0 ,\label{eq:susy_algebra}
\end{align}
which implies the conservation of $K$, $[H_Q,K] = 0$.  The Hamiltonian
decomposes into the eigenspaces of $K$ according to
\begin{align}
H_Q &= P_+ H_Q P_+ + P_- H_Q P_-, & P_\pm &= \tfrac12 (1 \pm K).
\end{align}
Since the Hamiltonian is given by the square of a Hermitian operator, the
eigenenergies are positive $E_n \geq 0$.  Given an eigenstate $|n,+\rangle$
from the ``bosonic'' sector with eigenvalue $E_n > 0$, the state $|n,-\rangle
= Q |n,+\rangle/\sqrt{E_n}$ is an eigenvector of the ``fermionic'' sector with
the same eigenvalue $E_n$.\cite{Note1}

Showing that our system \eqref{eq:ham_system} is supersymmetric amounts to
finding a supercharge $Q$ and an involution $K$ realizing the algebra
Eq.~\eqref{eq:susy_algebra}.  The first hunch that a potential supersymmetry
might be related to the parity of the actual number of electrons on the
superconducting island does not work due to the parity constraint
\eqref{eq:fermion_parity_constraint}, see App.~\ref{sec:app_susy}. However, the
Majorana Cooper-pair box has a ``hidden'' supersymmetry as we will show in the
following.

In order to show the supersymmetry of the Hamiltonian $H$ in the sense of
Eq.~\eqref{eq:susy_algebra}, we have to define an operator $K$
characterizing the sectors which commutes with the Hamiltonian.  For the
special case $n_g = 0$, such an operator is given by the parity $K:
\phi \mapsto -\phi$ of the superconducting phase difference.  We define the
supercharge\cite{Note2}
\begin{align}
Q &= \sqrt{E_C} \bigl[n - i \alpha \sin(\phi/2)\bigr] (-1)^n \\
&= \sqrt{E_C}  \{ n (-1)^n + \tfrac{i \alpha}2 [ (-1)^n \sin(\tfrac12\phi) -
  \sin(\tfrac12\phi)
(-1)^n] \}, \nonumber
\end{align}
where $\alpha$ is a free parameter and $(-1)^{n}$ is the fermion parity on the
superconducting island. It is straightforward to check that $[(-1)^n,
K]=0$ and $\{n - i \alpha \sin(\phi/2), K \}= 0$ such that
$Q$ anticommutes with $K$.

Using the trigonometric relation $2\sin^2(\phi/2) = 1-\cos\phi$, one obtains
the supersymmetric Hamiltonian 
\begin{align}
  H_Q = E_C \{ n^2  + \alpha \cos(\phi/2) +
    \tfrac12 \alpha^2 [ 1 - \cos(\phi)] \}
\end{align}
which for $\alpha = \sqrt{2 E_J/E_C}$ is equal to the Hamiltonian
\eqref{eq:ham_system} at the
point
\begin{align}\label{eq:susy_params}
  E_M &= \sqrt{2 E_J E_C}, & n_g &= 0.
\end{align}
The supersymmetry leads to a degeneracy of the spectrum (apart from the ground
state) which holds even in the nonperturbative regime of arbitrary $E_J/E_C$.
While the supersymmetry presented here does not depend on the presence of
fermionic degrees of freedom in the system, it relies crucially on the
presence of an anomalous Josephson junction in addition to a conventional
Josephson junction.  In this sense it provides a clear signature of the
Majorana-induced $4\pi$-periodic Josephson relation.

The supersymmetric structure of the Hamiltonian allows us to obtain the ground
state(s) to the energy eigenvalue zero by solving the first-order differential
equation $Q P_\pm \psi_0(\phi) = 0$ in the two sectors. In the present case,
there is no solution to $Q P_- \psi_0(\phi) = 0$ since solutions to $Q
\psi(\phi) = 0$ are always even with respect to the parity $K$ of the
superconducting phase difference. We obtain that the
non-degenerate ground state at the supersymmetric point 
Eq.~\eqref{eq:susy_params} is given by the function
\begin{equation}\label{eq:ground_state}
\psi_0(\phi) = [4 \pi I_0(2\alpha) ]^{-1/2} \exp[ - \alpha \cos (\phi/2) ]
\end{equation}
in the ``bosonic'' sector with the modified Bessel function $I_0(x)
=\int_0^{2\pi}\!dt\,\exp(x \cos t) /2\pi$.  All the excited states are doubly
degenerate due to the supersymmetry.  The algebraic construction of the higher
energy levels and states is unfortunately not possible, since the potential is
self-isospectral, that is identical in both sectors of the Hamiltonian, and
the algebraic construction works only for potentials that differ in at least
one parameter in the two sectors.\cite{cooper}

Despite the lack of a general analytic solution for the degenerate excited
states, we can still understand the degeneracy in the perturbative regimes.
In particular, in the limit $\alpha \to 0$ we recover the supersymmetry of the
free particle in one dimension as discussed by Ref.~\onlinecite{rau:04}.  In
this case the spectrum is given by $E_n = E_C n^2$ with the excess number of
electrons $n\in \mathbb{Z}$; the ground state corresponds to $n=0$, and the
level degeneracies are due to the two states $n$ and $-n$ at the same energy
for $n\geq 1$.\cite{Note3} It is an instructive exercise to check that the
level degeneracies persist when performing perturbation theory in $\alpha$.
What can be observed is that each term in order $N$ involving $E_J$ cancels
against a term in order $2N$ in $E_M$ appearing with opposite sign.  Thus, the
level degeneracy between $n$ and $-n$ persists to arbitrary order in $\alpha$,
see Fig.~\ref{fig:susy_wf}.  It is this particular cancellation of terms in
the perturbation theory for which supersymmetry as a non-perturbative
structure has been initially designed in the high-energy
context.\cite{weinberg}

In the semiclassical regime with $\alpha \to \infty$, the states are well
localized close to the minima at $\phi \in 2\pi \mathbb{Z}$ where the
potential $V=E_M \cos(\phi/2) + E_J (1-\cos \phi)$ can be expanded in
quadratic order, see Fig.~\ref{fig:susy_wf}(c).  Close to $\phi = 2\pi$, we
have $V_{2\pi} \approx -E_M + \tfrac12 E_J (\phi-2\pi)^2$ with the spectrum
$E_{2\pi,n} = - E_M + \sqrt{8 E_C E_J} (n+\tfrac12)$.  In the second mimimum,
at $\phi$ close to zero, we have $V_{0} \approx E_M + \tfrac12 E_J \phi^2$
which leads to the approximate spectrum $E_{0,n} = E_M + \sqrt{8 E_C E_J}
(n+\tfrac12)$.  At the supersymmetric point \eqref{eq:susy_params}, we observe
the degeneracy $E_{2\pi,n+1} = E_{0,n}$ valid for $n\geq 0$.  So the structure
is again a single ground state $E_{2\pi,0}$ with degenerate levels above it.
It is a highly nontrivial fact that the degeneracy found in the analysis above
valid for $\alpha \to \infty$ remains intact for finite $\alpha$ where next
order terms in the expansion of $V$ as well as tunneling events described by
instantons have to been taken into account.

In the following, we will show that the degeneracies of the whole spectrum
(except for the ground state) arising in the model \eqref{eq:ham_system} at
the supersymmetric point can be directly probed by a tunneling experiment.

\section{Tunneling current}\label{sec:tunneling_current}
\begin{figure*}[tbp]
\includegraphics[width=1.\textwidth]{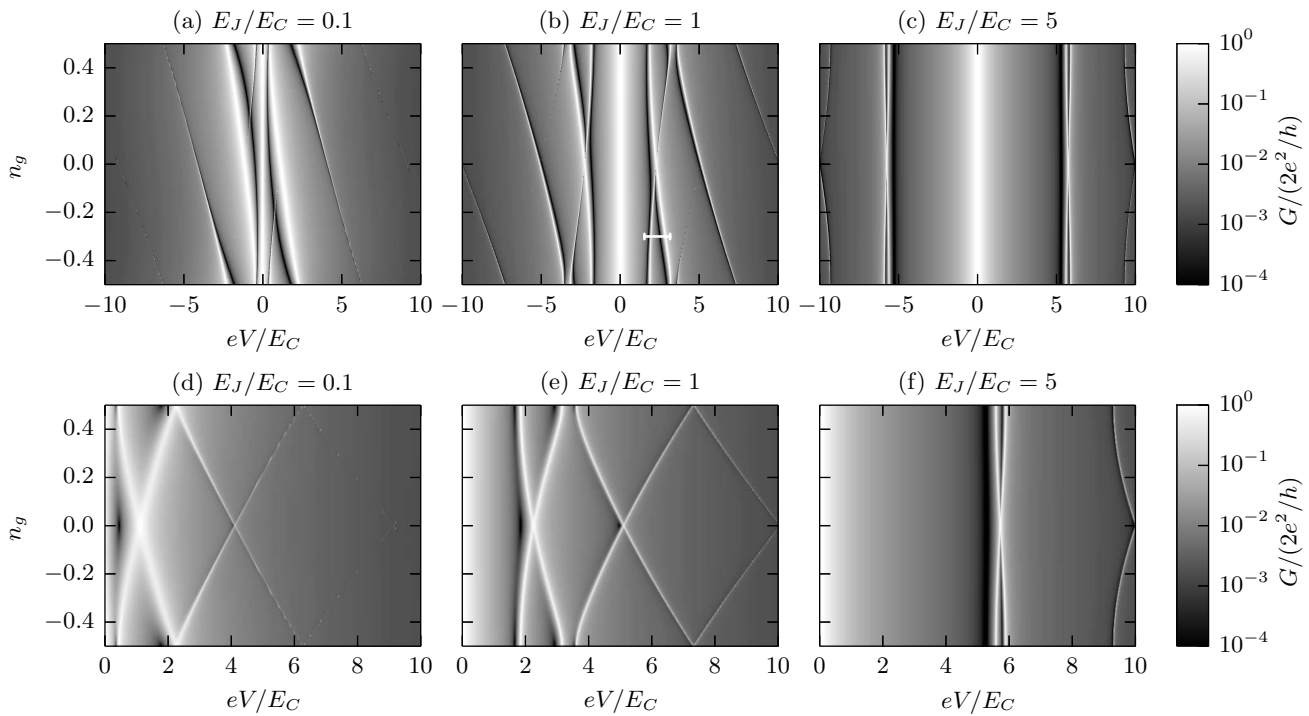}
\caption{The differential tunneling conductance $G$
Eq.~\eqref{eq:majorana_cond} at zero temperature at the special point $E_M =
\sqrt{2 E_J E_C}$ where the Hamiltonian of the system is supersymmetric for
$n_g=0$.  The conductances are plotted as a function of bias voltage $V$ and
offset charge $n_g$ for different ratios $E_J/E_C$ of Josephson energy energy
to the charging energy.  It is calculated from the exact bare Green's function
$G^R_{0,\varepsilon}$ of the system computed by exact diagonalization and
incorporating the leads via the Dyson equation~\eqref{eq:majorana_dyson} with
a tunnel coupling $\Gamma = 0.2 E_C$.  The upper panel displays the
conductance $G(V)$ discriminating between electron and hole processes.  The
strong bias asymmetry at low $E_J/E_C$ reflects the transition from the charge
basis to the phase basis as elaborated in the main text.  The lower panel
shows the symmetrized conductance $\tfrac12[G(V)+G(-V)]$ which is easier to
interpret.  In the symmetrized conductance, the crossing of all the levels at
$n_g=0$ is clearly visible. The white horizontal line in the upper panel (b)
between $eV/E_C = 1.2$ and $eV/E_C = 3.5$ at $n_g = -0.3$ indicates the position of the line cut shown in
Fig.~\ref{fig:fano_resonance}.}\label{fig:conductance_plots}
\end{figure*}

To model the tunneling experiment depicted in Fig.~\ref{fig:system_setup}, we
assume that the system is coupled via the Hamiltonian
\begin{align}
H_T = \sum_{p} w^* c_{p}^\dagger e^{-i \phi/2} \gamma_{1,B} + \text{H.c.}, \label{eq:ham_tunneling}
\end{align}
to an effective noninteracting lead of spinless electrons described by the
Hamiltonian $H_L = \sum_{p} \epsilon_{p} c_{p}^\dagger c_{p}$ with the
fermionic annihilation operators $c_p$;\cite{Note4} here, $w$ is the tunneling
matrix element and the presence of the operators $e^{\pm i \phi/2}$ account
for the transfer of charge to the superconductor.  For this setup, we derive
in the App.~\ref{sec:app_current} the exact expression
\begin{align} \label{eq:majorana_current}
I = \frac{e\Gamma}{h} \int\! d \varepsilon \,  (-\mathop{\rm Im} G^R_\varepsilon)
\left(2 f_{\varepsilon - eV} - 1\right)
\end{align}
for the tunneling current, see also Ref.~\onlinecite{hutzen:12}; here,
$f_\varepsilon = [1+\exp(\varepsilon/k_B T) ]^{-1}$ is the Fermi distribution
with respect to the chemical potential of the superconducting island and,
consequently, $f_{\varepsilon - eV}$ is the distribution of the electrons in
the lead.  The tunnel coupling $\Gamma = 2\pi |w|^2 \rho_0$ is given in the
wide-band limit where the electrons in the lead have the constant density of
state $\rho_0$.  The Majorana Green's function in the presence of the leads is
defined as
\begin{align}\label{eq:majorana_gf_retarded}
G^R_\varepsilon = -i \int_0^\infty \! d t \, e^{i \varepsilon t}  \big\langle \bigl\{\tilde \gamma(t),\, \tilde \gamma^\dagger \bigr\} \big\rangle,
\end{align}
with $\tilde \gamma = \gamma_{1,B} e^{-i \phi/2}$ evolving with respect to the
full Hamiltonian $H_\text{tot} = H_\gamma + H_L + H_T$.  From
Eq.~\eqref{eq:majorana_current}, the differential conductance $G(V) = d I/d V$
is obtained in the limit of low temperatures ($T\to 0$) where we can
approximate $d f_\varepsilon/d \varepsilon \approx -\delta(\varepsilon)$ as
\begin{align}\label{eq:majorana_cond}
  G(V) = \frac{d I}{d V} = -\frac{2e^2 \Gamma}h  \mathop{\rm Im} G^R_{eV}.
\end{align}
Our goal is to evaluate the differential conductances in the tunneling limit
$E_J, E_C \gg \Gamma$.  To this end, we need to relate the Majorana Green's
function $G^R_\varepsilon$ in presence of the leads to the Majorana Green's
function $G^R_{0,\varepsilon}$ without the leads ($w=0$) which can be
evaluated by exact diagonalization. 

In the non-interacting case ($E_C = 0$),
the retarded Majorana Green's function obeys a Dyson equation in Nambu
space,\cite{flensberg:10b} which can be written as
\begin{align} \label{eq:majorana_nambu_dyson}
	\check{G}^R_\varepsilon = \check{G}_{0,\varepsilon}^R + \check{G}_{0,\varepsilon}^R \check{\Sigma}^R_\varepsilon \check G^R_\varepsilon,
\end{align}
and where $\check{G}_\varepsilon$ is given by
\begin{align}\label{eq:majorana_nambu_gf}
  \check{G}^R_\varepsilon = -i \int_0^\infty \!\! d t \,e^{i\varepsilon t} \Bigl \langle\begin{pmatrix} \{ \tilde \gamma(t), \tilde \gamma^\dagger\}  & \{ \tilde \gamma(t), \tilde \gamma\}\\ 
		\{\tilde \gamma^\dagger(t), \tilde \gamma^\dagger\} & \{ \tilde \gamma^\dagger(t), \tilde \gamma\}
	\end{pmatrix}\Bigr \rangle,
\end{align}
and 
\begin{align}
 \check{\Sigma}^R_\varepsilon &= 
  |w|^2 \sum_p \begin{pmatrix} (\varepsilon - \epsilon_p +i 0^+)^{-1} & 0\\
  0 &  (\varepsilon + \epsilon_p +i 0^+)^{-1}  \end{pmatrix} \nonumber\\
  &=  - \frac{i \Gamma}2 \begin{pmatrix} 1 & 0 \\ 0 & 1 
 \end{pmatrix}
 \end{align}
is the self energy due to the lead.
In the non-interacting limit ($E_C = 0)$, the superconducting phase $\phi$ is
constant, $\phi(t) =\phi(0)$, and thus all the four entries of the Green's
function $\check G^R_\varepsilon$ are equal. As a consequence, the Dyson
equation becomes the scalar equation 
\begin{align}
G^R_\varepsilon = G_{0,\varepsilon}^R + G_{0,\varepsilon}^R \Sigma_\varepsilon^R G_\varepsilon^R\label{eq:majorana_dyson},
\end{align}
with the self energy $\Sigma^R_\varepsilon$ given by the sum of two processes
corresponding to transitions of electron and holes to the lead,
\begin{align}\label{eq:self_en}
  \Sigma^R_\varepsilon = \mathop{\rm tr} (\check\Sigma^R_\varepsilon )
  =
  -i \Gamma.
\end{align}

In the following, we assume that the scalar Dyson
equation~\eqref{eq:majorana_dyson} remains applicable also in the interacting
case.  This corresponds to a decoupling at the sequential tunneling level and
an inclusion of the leads through the self-energy \eqref{eq:self_en}.  We thus
neglect a potential difference in the dynamics between electrons and holes
when tunneling to the lead.\cite{Note5} As explained in the
App.~\ref{sec:app_bosonization_me}, the Green's function $G_{0,\varepsilon}^R$
at $w=0$ can be expressed in the Lehmann representation as
\begin{align}\label{eq:majorana_gf_lehmann}
  G^R_{0, \varepsilon} 
  = 
  \sum_{k,\sigma}
   \frac{a_k^\sigma}{\varepsilon - \sigma E_{k0} + i 0^+}
\end{align}
with the transition probabilities $a_k^\sigma = \vert \langle k \vert e^{i
\sigma \phi/2} \vert 0\rangle \vert^2$ to the exact eigenstate $| k\rangle$ of
the Hamiltonian $H$ by adding ($\sigma= +$) or removing ($\sigma =-$) a single
electron, where $E_{kl} = E_k -E_l$ are differences of the corresponding
eigenenergies.  The full retarded Green's function $G^R_\varepsilon$ is
obtained via the Dyson equation~\eqref{eq:majorana_dyson}.  The effect of the
leads incorporated via the Dyson equation~\eqref{eq:majorana_dyson} is to
provide a state-dependent broadening of the levels of the isolated system.

To get a feeling for the formulas, we first consider the simple situation
where due to a large level separation only a single level $| k \rangle$ is
close to resonance $eV \approx E_{k0} >0$.  In this case, we can approximate
$G_{0,\varepsilon}^R \approx a_k^+/(\varepsilon - E_{k0} + i 0^+)$.  Resolving
the Dyson equation and plugging the resulting expresssion for
$G_\varepsilon^R$ into \eqref{eq:majorana_cond} yields
\begin{align}
  G(V) \approx \frac{2 e^2}h 
  \frac{(a_k^+ \Gamma)^2}{(eV - E_{k0})^2 + (a_k^+
	\Gamma)^2},\label{eq:cond_single_level}
\end{align}
which describes a Lorentzian peak around the resonance energy $E_{k0}$ with
level-dependent broadening $a_k^+ \Gamma$ proportional to the probability of
injecting an electron from the lead. 

As a peak in the conductance is associated with the resonance condition $eV=
E_{k0}$, we expect that the level degeneracy due to the supersymmetry is
visible as a merging of two peaks when approaching the supersymmetric point.
To test this hypothesis, we have numerically calculated the conductance by
determining $G^R_{0,\varepsilon}$ via exact diagonalization of $H$ and
subsequently resolving the Dyson equation \eqref{eq:majorana_dyson}.  The
resulting differential conductance Eq.~\eqref{eq:majorana_cond} is displayed
in Fig.~\ref{fig:conductance_plots} as a function of bias voltage and offset
charge $n_g$.  As was to be expected from the approximate expression
Eq.~\eqref{eq:cond_single_level}, the conductance peaks with a value equal to
the conductance quantum $2e^2/h$ when the bias voltage is tuned such that the
chemical potential of the lead is in resonance with the eigenstates of the
isolated system and resonant Andreev reflection occurs. The conductance plots
exhibit the symmetry $G(V, n_g) = G(-V,-n_g)$ which is exact to our
level of approximation. For the Hamiltonian $H$, a sign flip of the  offset
charge $n_g$ is equivalent to a sign flip of the superconducting phase under
action of the parity operator $K: \phi \mapsto -\phi$.
It is easily checked from the Lehmann representation
Eq.~\eqref{eq:majorana_gf_lehmann} that $G_{0,\varepsilon}^R \mapsto -
G_{0,-\varepsilon}^A$ under the operation of $K$ which due to the strucure of
the Dyson equation~\eqref{eq:majorana_dyson} translates into
$G(V) \mapsto G(-V)$. An intuitive reason 
for the symmetry $G(V, n_g) = G(-V,-n_g)$ is that the sign of $n_g$ favors an excess or defect number of electrons 
whereas the sign of $V$ corresponds to the lead preferably adding or removing
electrons.
As we have shown in Sec.~\ref{sec:susy_qm_cpb}, the supersymmetry at small 
$E_J/E_C$ corresponds to a different sign of excess electrons.  Thus the two
levels that cross appear in the tunneling conductance at opposite bias.  In
order to remedy the problem that the crossing is not directly visible in the
conductance as it appears at different bias, we plot the symmetrized
conductance $\tfrac12[G(V) + G(-V)]$ in the lower panel of 
Fig.~\ref{fig:conductance_plots} where the crossing of the
levels at $n_g=0$ is visible for all ratios of $E_J$ to $E_C$.  Due to the
symmetry mentioned above, the plots of the symmetrized conductances are
symmetric under $n_g \leftrightarrow -n_g$.

\begin{figure}[tbp]
	\includegraphics[width=.9\columnwidth]{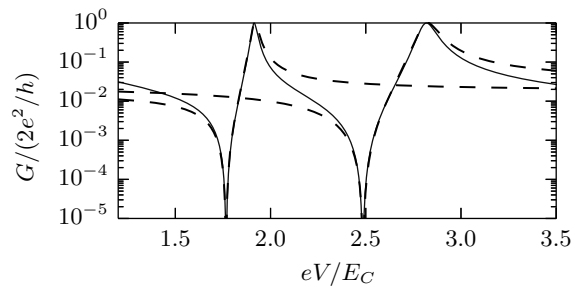}
  \caption{\emph{Solid line:} Line cut through the tunneling conductance
		$G(V)$ shown in Fig.~\ref{fig:conductance_plots} for $E_J/E_C
		= 1$, $E_M = \sqrt{2 E_J E_C}$, $\Gamma = 0.2 E_C$ and $n_g =
		-0.3$ as a function of bias voltage $V$.  \emph{Dashed lines:}
		fits of Fano peaks of the form
		Eq.~\eqref{eq:fano_resonance_form} with parameters $\beta =
		7.5$, $\gamma/E_C \approx 0.04$, $\varepsilon_0/E_C \approx
		1.91$ for the resonance on the left and $\beta = 6$,
		$\gamma/E_C \approx 0.11$, $\omega_0/E_C \approx 2.81$ for the
		resonance on the right.}.\label{fig:fano_resonance}
\end{figure}

A striking feature which is especially visible in the unsymmetrized
conductance plots is the coincidence of peaks in the conductances with regions
of suppressed conductance.  In Fig.~\ref{fig:fano_resonance}, we have shown a
linecut of the conductance as a function bias voltage $V$ at the point
$E_J/E_C = 1$, $E_M = \sqrt{2 E_J E_C}$ and $n_g = -0.3$ for $\Gamma = 0.2
E_C$.  It shows a sequence of two conductance peaks with their associated
minima to the left of them. Each of the conductance maxima-minima pair  has
been fitted to a Fano resonances of the form
\begin{align} 
  G_\text{Fano}(V) = \frac{2 e^2}h \frac{ (
  \beta \gamma /2 + \varepsilon - \varepsilon_0)^2}{
    (1+\beta^2) [  (\varepsilon -
    \varepsilon_0)^2 + \gamma^2/4 ]}\label{eq:fano_resonance_form}
\end{align}
indicated by the dashed lines; here, $\beta \in \mathbb{R}$ is the asymmetry
parameter, $\gamma$ is the line-width of the resonance, and $\varepsilon_0$
the position of the resonance.  In the limit $|\beta|\rightarrow \infty$, the
Fano resonance approaches the usual Breit-Wigner resonance.  It is clear from
Fig.~\ref{fig:fano_resonance} that the Fano-resonance behavior captures the
behavior of the conductance close to the maximum.

In order to understand the microscopic origin of the Fano-like resonances, we
study a simplified model with $E_M = 0$ in the regime $E_C \gg E_J \gg
\Gamma$.  In this case, we can evaluate the Hamiltonian
Eq.~\eqref{eq:ham_system} in the charge basis and truncate to charge states
$|n=0\rangle, |n=\pm 1 \rangle$ which yields the effective three-level
Hamiltonian
\begin{align}\label{eq:ham_truncated} 
H \approx E_J + \begin{pmatrix} E_C(1 + n_g)^2 & 0 & -E_J/2 \\ 0
	& E_C n_g^2 & 0 \\ - E_J/2 & 0 & E_C(1 - n_g)^2
\end{pmatrix} 
\end{align}
with ground state $|0\rangle = |n=0\rangle$ at eigenenergy $E_0 = E_J+E_C
n_g^2$ (which is exact for all $E_J/E_C$ and $n_g$ due to the fact that we
have set $E_M=0$) and excited states $| 1,2 \rangle$ with eigenenergies
$E_{1,2}$.

Due to the algebra Eq.~\eqref{eq:charge_phase_algebra}, we find $a^\sigma_0 =
|\langle 0 | e^{\sigma i \phi/2} | 0 \rangle|^2 = 0$.  The completeness
relation leads to $a_1^\sigma + a_2^\sigma = 1$.  At the particle-hole
symmetric point $n_g=0$, we have that $a_1^\sigma = a_2^\sigma = \tfrac12$.
For $n_g \neq 0$, the particle-hole symmetry is broken and thus $a_1^\sigma
\neq a_2^\sigma$.  The bare Green's function $G_{0,\varepsilon}^R$ of the
effective three-level system follows from the Lehmann representation
Eq.~\eqref{eq:majorana_gf_lehmann}.  We consider the case of $V>0$ such that
the main contribution arises from the terms with $\sigma =+$. Solving the
Dyson equation for $G^R_\omega$, we obtain the expression
\begin{align} \label{eq:cond_two_level}
  G(V) \approx \frac{2 e^2}h \frac{(x - a_1^+)^2}{(x-a_1^+)^2 + E_{21}^2 x^2 
		(1-x)^2/\Gamma^2} 
\end{align}
for the conductance, where we have replaced the voltage by the dimensionless
variable $x = (eV - E_{10})/E_{21}$ that is centered around the resonance at
$eV=E_{10}$.\cite{Note6} Note that in the limit of large level-separation $x
\to 0$, we recover the single-level conductance
Eq.~\eqref{eq:cond_single_level}.

\begin{figure}[tbp]
	\includegraphics[width=.9\columnwidth]{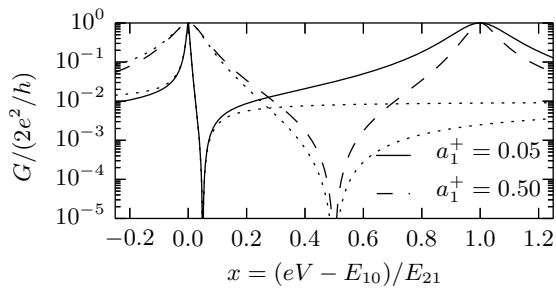}
	\caption{Differential tunneling conductance
		Eq.~\eqref{eq:cond_two_level} into the effective three-level
		system Eq.~\eqref{eq:ham_truncated} for $E_{21}/\Gamma = 10$
		and different values of the
		transition probability $a_1^+$ into the state $| 1 \rangle$. Dotted lines
		in the vicinity to the actual curves show the approximate
		Fano-resonance form Eq.~\eqref{eq:fano_resonance_form} valid
		for $x = (\varepsilon - E_{10})/E_{21} \ll 1$ that is parametrized by 
		$\beta = -E_{21}/\Gamma$, $\varepsilon_0 = E_{10} + a_1^+
		E_{21}/(1+\beta^2)$ and $\gamma = 2 a_1^+ \Gamma \beta^2/(1+\beta^2)$.} \label{fig:cond_fano_model}
\end{figure}

The new feature brought by the inclusion of the second level is a zero in the
conductance for $\varepsilon$ between $E_{10}$ and $E_{20}$ at the position $x
= a^+_1$.\cite{Note7} The zero in the conductance arises due to the
competition of the processes of tunneling an electron into level $|1\rangle$
and level $|2\rangle$ leading to an interference.  The interference can be
traced back to the fact that the process happens at an energy $eV > E_{10}$
which is above the resonance at $E_{10}$ and thus is approximately phase
shifted by $\pi$ with respect to the second level where $eV < E_{20}$.  To see
that the expression \eqref{eq:cond_two_level} of the differential conductance
is of the Fano form close to the resonance at $x=0$, we expand $x$ to second
order in the denominator and obtain $G_\text{Fano}(V)$ for the conductance
around the resonance $E_{10}$, where $-\beta = E_{21}/\Gamma \gg 1$, $\gamma
\approx 2 a^+_1 \Gamma$, and $\omega_0 \approx E_{k0}$.  Note that the
asymmetry parameter $\beta$ is negative in accord with the fact that the root
in the conductance occurs to the right of the resonance position $E_{10}$ and
that $\gamma$ and $\omega_0$ fit the single level result
\eqref{eq:cond_single_level}.  Since $\beta \gg 1$ in the tunneling regime,
the zero is at a position where the conductance is already polynomially
suppressed, but the dip in the conductance still clearly shows up in a
logarithmic scale, cf.\ Fig.~\ref{fig:fano_resonance}.  For $E_C \gg E_J$, we
have $a^+_2 \rightarrow 0$ and we find a very sharp resonance at $E_{20}$ with
dip in the conductance in close vicinity to $E_{20}$ (as compared to the peak
separation $E_{21}$) whereas the resonance at $E_{10}$ assumes its usual
Breit-Wigner form, see Fig.~\ref{fig:cond_fano_model}.  It is interesting to
note that similar interference effects are known from transport through
molecules when multiple transport channels are available.\cite{guedon:12,
geranton:13}

\section{Conclusions}

We have shown that the extension of the usual Cooper-pair box by a
Majorana-Josephson junction features a degeneracy in its spectrum for $n_g =
0$ and $E_M = \sqrt{2 E_J E_C}$ that is due to a ``hidden'' bosonic
supersymmetry generalizing the supersymmetry of the free particle.  The
supersymmetry crucially relies on the presence of the anomalous Josephson
junction and an observation of the predicted level crossings of all excited
states at the supersymmetric point provides a clear indication of the presence
of a Majorana-induced anomalous Josephson junction. 

We have shown that the
supersymmetry can be probed directly in a tunneling experiment by varying the
bias voltage $V$ and the offset charge $n_g$.  In the tunneling regime, when
the bias voltage is tuned such that the Fermi level of the lead coincides with
the eigenenergies of the isolated system, resonant Andreev reflection with a
peak conductance equal to the conductance quantum $2 e^2/h$ occurs for the
whole range of $E_J/E_C$ values.  The crossing of all excited eigenstates of
the system as the supersymmetric point is approached can thus clearly be
observed in conductance maps obtained by varying the offset charge $n_g$ and
gate voltage $V$.  The conductance features suppressed conductance close to
the resonances that are due to interference.  We have explained that the
interference is due the presence of several channels for single-electron
tunneling which exist since the island charge is not a sharp observable by
considering a simple analytic model in the regime $E_J/E_C \ll 1$.

Interestingly, the supersymmetry presented here could also be found in
Josephson Rhombi chains allowing tunneling of pairs of Cooper
pairs\cite{doucot:02} in addition to conventional Cooper-pair tunneling that
are constructed in conventional Josephson junction arrays and have recently
been realized experimentally.\cite{bell:13} However, in this case the Hilbert
space only involves the superconducting condensate and the states and potential
degeneracies of the system cannot be simply probed by tunneling spectroscopy as
proposed in this paper.  It is an interesting question for future work, if
there is a simple experimental signature of supersymmetry in this setup.

\begin{acknowledgments}
 We acknowledge fruitful discussions with C.-Y.  Hou and A. M. Tsvelik. JU and
 FH are grateful for support from the Alexander von Humboldt foundation and
 from the RWTH Aachen University Seed Funds. DS acknowledges support of the
 D-ITP consortium, a program of the Netherlands Organisation for Scientific
 Research (NWO) that is funded by the Dutch Ministry of Education, Culture and
 Science (OCW).
\end{acknowledgments}

\appendix
\section{Bosonization of the Hamiltonian}
\label{sec:app_bosonization_ham}
Bosonizing the Majorana operators of the Hamiltonian Eq.~\eqref{eq:ham_system_general} via a Jordan-Wigner transformation as
\begin{align}\label{eq:jw_spin_correspondance}
\gamma_{k,A} &= \Big( \prod_{l < k} \, \sigma_l^z \Big)  \sigma_k^x &
\gamma_{k,B} &= -\Big( \prod_{l < k} \, \sigma_l^z \Big) \sigma_k^y,
\end{align}
where $\sigma_k^{x,y,z}$ are independent sets of Pauli-matrices for each index $k$, 
brings the Majorana tunneling term to the form
\begin{align}
E_M \sigma_0^x \sigma_1^x \cos(\phi/2) = E_M \cos((\phi + \pi \sigma_0^x - \pi
\sigma_1^x)/2).
\end{align}
Performing a unitary transformation $U = \prod_{k=0,1} U_k$ with
\begin{align}\label{eq:spin_unitary_op}
U_k = e^{-i \pi \sigma_k^x n_k/2} = \cos(\pi n_k/2) - i \sigma_k^x \sin(\pi
n_k/2),
\end{align}
which links the transfer of one electron to a flip of the fermion parity and
where we use the notation $n_1 = n$, the transformed Hamiltonian assumes the form
\begin{align}
U H_\gamma U^\dagger = \Big\{ E_C (n - n_g)^2 
+ E_J (1 - \cos \phi) \nonumber \\
+ E_M \cos(\phi/2) \Big\} I, \label{eq:ham_bosonized}
\end{align}
while the fermion parity constraint Eq.~\eqref{eq:fermion_parity_constraint} is transformed into
\begin{align}
\cos(\pi n_k) \sigma_k^z - \sin(\pi n_k) \sigma_k^y = (-1)^{n_k} .
\end{align}
In Eq.~\eqref{eq:ham_bosonized}, we have made explicit the trivial remaining
spin-structure of the Hamiltonian. On the other hand, the transformed
constraint fixes integer charges $n_k \in \mathbb{Z}$ and enforces spin-up
eigenstates of $\sigma_k^z$. The fermion-parity constraint is thus resolved in
the transformed Hamiltonian Eq.~\eqref{eq:ham_bosonized} by considering just
one spin-component and demanding a $4\pi$ periodicity of the eigenstates,
leading to the Hamiltonian Eq.~\eqref{eq:ham_system} given in the main text.

\section{Supersymmetry without the fermion parity constraint}
\label{sec:app_susy}

In a theory not bound by the fermion parity constraint
Eq.~\eqref{eq:fermion_parity_constraint}, the fermion parity $K^\prime = i
\gamma_{0,B} \gamma_{1,A}$ across the Majorana junction is easily seen to be
conserved in the original Hamiltonian $H_\gamma$ from
Eq.~\eqref{eq:ham_system_general}. The natural choice for the supercharge 
$Q^\prime$ is then given by
\begin{align}
Q^\prime = \sqrt{E_C} \bigl[(n - n_g) \, \gamma_{0,B} - \alpha^\prime
\sin(\phi/2) \gamma_{1,A}\bigr] \label{eq:susy_wrong},
\end{align}
where $\alpha^\prime$ is a free parameter. The charge $Q^\prime$ anticommutes
with the fermion parity $K^\prime$ across the junction. One verifies that $Q'^2
= H_\gamma$ for the parameters
Eq.~\eqref{eq:susy_params} given in the main text and $\alpha^\prime = \alpha = \sqrt{2 E_J/E_C}$.
The supersymmetry described above corresponds to a supersymmetry between
bosonic and fermionic sectors, which cannot be realized in our system since the
Majoranas are no longer an independent degree of freedom due to the fermion
parity constraint Eq.~\eqref{eq:fermion_parity_constraint} and only the
bosonized ``hidden'' supersymmetry given in the main text remains.

\section{Derivation of the current}
\label{sec:app_current}
Let us define a generic tunneling Hamiltonian
\begin{align}
H_T = \sum_{ilp} w_{lpi}^* c_{lp}^\dagger e^{-i \phi_i/2} \gamma_k + w_{lpi}
\gamma_i e^{i \phi_i/2} c_{lp},
\end{align}
describing the tunneling with tunneling matrix elements $w_{lpi}$ between electrons of momentum $p$ in lead $l$ with creation/annihilation operators $c_{lp}^\dagger, c_{lp}$ into Majorana bound states $\gamma_i$ with associated superconducting phases $\phi_i$ ($\phi_i = \phi_j$ for Majoranas on the same superconductor).
The leads are free with Hamiltonian
\begin{align}
H_L = \sum_l \sum_p \epsilon_{lp} c_{lp}^\dagger c_{lp}.
\end{align}
Defining operators $\tilde \gamma_i = e^{-i \phi_i/2} \gamma_i$, the tunneling Hamiltonian has the appearance of a standard fermionic tunneling Hamiltonian. 
It is well-known \cite{[{}][{, Chap. 12.4; }]haug, *[][{, Chap. 8.9}]datta}
that the expression for the steady-state current $I_l$ through lead $l$, given
a system with a tunneling Hamiltonian of the form Eq.~\eqref{eq:ham_tunneling} and non-interacting leads, can always be cast in the form 
\begin{align}
  I_l = \frac{e}{\hbar} \int \frac{d \varepsilon}{2\pi} \mathop{\rm tr} \big[ (i G_\varepsilon^>) \Gamma_\varepsilon^l f^l_\varepsilon - (-i G_\varepsilon^<) \Gamma_\varepsilon^l(1-f_\varepsilon^l)\big],\label{eq:majorana_current_intermediate}
\end{align}
where only the non-interaction of the leads has been exploited and 
\begin{align}
(G^<_{\varepsilon})_{ij} &= i \int d t \, e^{i \varepsilon t} \, \big\langle \tilde \gamma_{j}^\dagger \tilde \gamma_i(t) \big\rangle \\
(G^>_{\varepsilon})_{ij} &= -i \int d t \, e^{i \varepsilon t} \, \big\langle \tilde \gamma_i(t) \tilde \gamma_{j}^\dagger \big\rangle
\end{align}
are the lesser/greater Majorana Green's functions in presence of the leads in
the steady state limit, $f^l_{\varepsilon}$ is the equilibrium Fermi
distribution of lead $l$ and $(\Gamma_\varepsilon^l)_{ij} = 2 \pi \sum_{p}
w_{lpi} w_{lpj}^* \delta(\varepsilon - \epsilon_{lp})$ is the lead coupling
matrix. The current expression Eq.~\eqref{eq:majorana_current_intermediate} has a straight-forward interpretation: the contribution to the current through lead $l$ at energy $\varepsilon$ is given by the rate  $\Gamma_\varepsilon^l f_\varepsilon^l$ of electron tunneling into the system through lead $l$ at energy $\varepsilon$ times the number $i G^>_\varepsilon$ of available states at this energy  minus the rate $\Gamma_\varepsilon^l (1-f_\varepsilon^l)$ of electrons tunneling out of the system into lead $l$ times the number $-i G^<_\varepsilon$ of occupied states.\cite{[][{, Chap. 8.9}]datta}
With the Keldysh Green's function $G^K = G^> + G^<$ and the relation $G^R - G^A
= G^> - G^<$, one can rewrite the expression Eq.~\eqref{eq:majorana_current_intermediate} in the form
\begin{align}
  I_l &= \frac{e}{\hbar} \int \frac{d \varepsilon}{2\pi}  \mathop{\rm tr} \Big\{ \big[i (G_\varepsilon^R - G_\varepsilon^A) + i G^K_\varepsilon\big]\Gamma_\varepsilon^{l} f_\varepsilon^l \nonumber  \\  
& \qquad - \big[i (G_\varepsilon^R - G_\varepsilon^A) - i G^K_\varepsilon\big]\Gamma_\varepsilon^{l} (1-f_\varepsilon^l) \Big\} \label{eq:appendix_current_intermediate}\\
&= \frac{e}{\hbar} \int \frac{d \varepsilon}{2\pi} \mathop{\rm tr} \Big\{i (G_\varepsilon^R - G_\varepsilon^A) \Gamma_\varepsilon^l (2 f^l_\varepsilon - 1) \nonumber \\
& \qquad + \frac{i}{2} G^K_\varepsilon \Gamma_\varepsilon^l\Big\}.
\end{align}
The key observation from Ref.~\onlinecite{hutzen:12} is that the $G_\varepsilon^K$ expression vanishes when working with a wide band, $\Gamma^l_\varepsilon = \Gamma^l$, and assuming coupling to just one Majorana $\gamma_k$, $(\Gamma^l)_{ij} = (\Gamma^l)_{kk} \delta_{ik} \delta_{jk}$, i.e.,
\begin{align}
0 = \int \frac{d \varepsilon}{2\pi} \mathop{\rm tr}[G_\varepsilon^K \Gamma_\varepsilon^l] = (\Gamma^l)_{kk} G^K(t=0)_{kk}
\end{align}
since $G^K(t=0)_{kk} =  -i \big\langle [ \tilde \gamma_k, \tilde \gamma_k^\dagger]\big\rangle = 0$. Thus, in the wide-band limit with coupling to just one Majorana $\gamma_k$, one obtains the expression
\begin{align}
I_l = \frac{e}{\hbar} \int \frac{d \varepsilon}{2\pi} \big(-\mathop{\rm Im} G_\varepsilon^R\big)_{kk} (\Gamma^l)_{kk} (2 f^l_\varepsilon - 1),
\end{align}
valid both in non-interacting and interacting setups. The very cumbersome feature brought by the Majoranas is that even in the interacting case the current expression remains completely independent from the occupation state of the system and depends only on spectral properties. This can be seen as yet another reflection of the fact that a single Majorana mode does not have a well-defined occupation number. 

\section{Evaluation of transition probabilities}
\label{sec:app_bosonization_me}
The eigenstates $|k_\gamma\rangle$ of the Hamiltonian
$H_\gamma$ from Eq.~\eqref{eq:ham_system_general} are related to the eigenstates
$|k\rangle$ of the bosonized Hamiltonian $H$ from Eq.~\eqref{eq:ham_system} via
the unitary transformation Eq.~\eqref{eq:spin_unitary_op} as $|n\rangle
= U |n_\gamma\rangle$. The Majorana $\gamma_{1,B}$ is in the bosonized form
expressed as $\gamma_{1,B} = -\sigma_0^z \sigma_1^y$. Using $U \sigma_k^y U^\dagger = (-1)^{n_k} \sigma_k^y$,
$U \sigma_k^z U^\dagger = (-1)^{n_k} \sigma_k^z$ and $U e^{\pm i \phi_k/2}
U^\dagger = \mp i \sigma_k^x e^{\pm i \sigma_k^x/2}$, where we again identify
$n_1 = n$, one finds
\begin{align}
U \gamma_{1,B} e^{\pm i \phi/2} U^\dagger = \mp \sigma_0^z \sigma_1^z e^{\pm i
	\phi/2} (-1)^{n_0 + n}.
\end{align}
Since $n_0 + n$ and $\sigma_k^z$ are conserved quantities of the
Hamiltonian Eq.~\eqref{eq:ham_bosonized}, one obtains for the transition
probabilites
$|\langle k_\gamma | \gamma_{1,B} e^{\pm i \phi/2} | l_\gamma \rangle|^2$ the result
\begin{align}
  |\langle k_\gamma | \gamma_{1,B} e^{\pm i \phi/2} | l_\gamma \rangle|^2 = |
  \langle k | e^{\pm i \phi/2} | l \rangle |^2
\end{align}
employed in the main text.


\begin{thebibliography}{10}
\bibitem{witten:81}
E. Witten,
 Nucl. Phys. B {\bf 188} (3), 513  (1981).

\bibitem{cooper:83}
F. Cooper and B. Freedman,
 Ann. Phys. (NY) {\bf 146} (2), 262  (1983).

\bibitem{cooper:95}
F. Cooper, A. Khare, and U. Sukhatme,
 Phys. Rep. {\bf 251} (5-6), 267 (1995).

\bibitem{fendley:03b}
P. Fendley, B. Nienhuis, and K. Schoutens,
 J. Phys. A {\bf 36}, 12399 (2003).

\bibitem{fendley:03}
P. Fendley, K. Schoutens, and J. de~Boer,
 Phys. Rev. Lett. {\bf 90}, 120402 (2003).

\bibitem{beccaria:05}
M. Beccaria and G. F. De~Angelis,
 Phys. Rev. Lett. {\bf 94}, 100401 (2005).

\bibitem{huijse:08}
L. Huijse, J. Halverson, P. Fendley, and K. Schoutens,
 Phys. Rev. Lett. {\bf 101}, 146406 (2008).

\bibitem{tangerman:93}
R. D. Tangerman and J. A. Tjon,
 Phys. Rev. A {\bf 48}, 1089 (1993).

\bibitem{snoek:05}
M. Snoek, M. Haque, S. Vandoren, and H. T. C. Stoof,
 Phys. Rev. Lett. {\bf 95}, 250401 (2005).

\bibitem{yu:08}
Y. Yu and K. Yang,
 Phys. Rev. Lett. {\bf 100}, 090404 (2008).

\bibitem{plyushchay:96}
M. S. Plyushchay,
 Mod. Phys. Lett. A {\bf 11}, 397 (1996).

\bibitem{rau:04}
A. R. P. Rau,
 J. Phys. A {\bf 37}, 10421 (2004).

\bibitem{alicea:12}
J. Alicea,
 Rep. Prog. Phys. {\bf 75}, 076501 (2012).

\bibitem{beenakker:13}
C. W. J. Beenakker,
 Annu. Rev. Con. Mat. Phys. {\bf 4}, 113 (2013).

\bibitem{nayak:08}
C. Nayak, S. H. Simon, A. Stern, M. Freedman, and S. Das~Sarma,
 Rev. Mod. Phys. {\bf 80}, 1083 (2008).

\bibitem{tsvelik:12}
A. M. Tsvelik,
 Europhys. Lett. {\bf 97}, 17011 (2012).

\bibitem{grover:13}
T. Grover, D. N. Sheng, and A. Vishwanath,
 arXiv:1301.7449 (2013).

\bibitem{oreg:10}
Y. Oreg, G. Refael, and F. von Oppen,
 Phys. Rev. Lett. {\bf 105}, 177002 (2010).

\bibitem{lutchyn:10}
R. M. Lutchyn, J. D. Sau, and S. Das~Sarma,
 Phys. Rev. Lett. {\bf 105}, 077001 (2010).

\bibitem{fu:10}
L. Fu,
 Phys. Rev. Lett. {\bf 104}, 056402 (2010).

\bibitem{heck:11}
B. van Heck, F. Hassler, A. R. Akhmerov, and C. W. J. Beenakker,
 Phys. Rev. B {\bf 84}, 180502(R) (2011).

\bibitem{zazunov:11}
A. Zazunov, A. L. Yeyati, and R. Egger,
 Phys. Rev. B {\bf 84}, 165440 (2011).

\bibitem{combescure:04}
M. Combescure, F. Gieres, and M. Kibler,
 J. Phys. A {\bf 37}, 10385 (2004).

\bibitem{Note1}
Note that only when the sign of $K$ corresponds to the parity of physical
  fermions in the system, the resulting structure is a true supersymmetry
  between bosons and fermions. In general, the symmetry of the Hamiltonian is
  not in any way related to actual fermions or bosons present in the system.

\bibitem{Note2}
In going from the first to the second line, we have used the fact that $(-1)^n$
  anticommutes with $\sin(\phi /2)$ as the latter
  creates transitions between different fermion parity states.

\bibitem{cooper}
F. Cooper, A. Khare, and U. Sukhatme,
 {\em Supersymmetry in Quantum Mechanics\/}
 (World Scientific, 2001).

\bibitem{Note3}
We obtain the free particle spectrum by the relation $E_C = \hbar ^2/2m L^2$
  with $m$ the mass of the particle and $L$ the length of of the system.

\bibitem{weinberg}
S. Weinberg,
 {\em The Quantum Theory of Fields: Supersymmetry\/},
 The Quantum Theory of Fields (Cambridge University Press, 2000).

\bibitem{Note4}
If the leads are spin-degenerate, only one of the Kramers' partners is coupled
  to the Majorana mode while the other is completely reflected and thus our
  results remain valid in this case.

\bibitem{hutzen:12}
R. H{\"u}tzen, A. Zazunov, B. Braunecker, A. L. Yeyati, and R. Egger,
 Phys. Rev. Lett. {\bf 109}, 166403 (2012).

\bibitem{flensberg:10b}
K. Flensberg,
 Phys. Rev. B {\bf 82}, 180516(R) (2010).

\bibitem{Note5} Alternatively, one could work with the Nambu-Dyson
equation~\eqref{eq:majorana_nambu_dyson}, which contains the
  different dynamics through the anomalous Green's functions of type $\langle
  \tilde{\gamma} \tilde{\gamma} \rangle$, $\langle \tilde{\gamma}^\dagger
  \tilde{\gamma}^\dag \rangle$.  We have checked that this does not
  qualitatively alter the results and thus we proceed with the simpler
  approach, i.e., the decoupling given in the main text.

\bibitem{Note6}
The conductance can also be expressed as a function of $1-x = (E_{20}
  -\varepsilon )/E_{21}$ centered around the second resonance at $E_{20}$
  through the symmetry $a_1^+ \leftrightarrow a_2^+= 1-a_1^+$ and $x
  \leftrightarrow 1-x$ which can be used for the analysis of the expression
  close to $E_{20}$.

\bibitem{Note7}
That the conductance is exactly zero is an artifact of the restricted Hilbert
  space. In general, the zero is replaced by a suppression of the conductance.

\bibitem{guedon:12}
C. M. Gu{\'e}don, H. Valkenier, T. Markussen, K. S. Thygesen, J. C. Hummelen,
  and S. J. van~der Molen,
 Nat. Nanotech. {\bf 7} (5), 305 (2012).

\bibitem{geranton:13}
G. G\'{e}ranton, C. Seiler, A. Bagrets, L. Venkataraman, and F. Evers,
 J. Chem. Phys. {\bf 139} (23), 234701 (2013).

\bibitem{doucot:02}
B. Doucot and J. Vidal,
 Phys. Rev. Lett. {\bf 88}, 227005 (2002).

\bibitem{bell:13}
M. T. Bell, J. Paramanandam, L. B. Ioffe, and M. E. Gershenson,
 arXiv:1311.6521 (2013).

\bibitem{haug}
H. Haug and A. Jauho,
 {\em Quantum Kinetics in Transport and Optics of Semiconductors\/},
 Solid-State Sciences (Springer, 2007).

\bibitem{datta}
S. Datta,
 {\em Electronic Transport in Mesoscopic Systems\/}
 (Cambridge University Press, Cambridge, 1995).

\end{thebibliography}
\end{document}